\documentclass[aps,prb,twocolumn,superscriptaddress,showpacs]{revtex4-1}

\usepackage{hyperref}
\usepackage[dvips]{graphicx}
\usepackage{amssymb,amsmath,amsfonts}
\usepackage[T1]{fontenc}
\usepackage{color}
\usepackage{changes}
\newcommand{\rethree}{$(\sqrt{7}\times\sqrt{7})R19.1^{\circ}$}

\newcommand{\retwo}{$(\sqrt{3}\times\sqrt{3})R30^{\circ}$}
\newcommand{\ke}{($\vec{k}, E$) }
\newcommand{\mke}{$\vec m(\vec k ,E)$}
\newcommand{\mkk}{$\vec m(\vec k)$}
\newcommand{\mpar}{$m_\parallel$}
\newcommand{\mper}{$m_\perp$}

\newcommand{\up}{$\uparrow$}
\newcommand{\down}{$\downarrow$}

\begin{document}


\title{Complex spin texture of Dirac cones induced via spin-orbit proximity effect in graphene on metals}


\author{Jagoda S\l awi\'{n}ska*}
\affiliation{Instituto de Ciencia de Materiales de Madrid, ICMM-CSIC, Cantoblanco, 28049 Madrid, Spain}
\affiliation{Department of Solid State Physics, University of \L \'{o}d\'{z}, Pomorska 149/153, 90236 \L \'{o}d\'{z}, Poland}
\affiliation{Consiglio Nazionale delle Ricerche, Istituto SPIN L'Aquila, sede temporanea di Chieti, 66100 Italy}
\altaffiliation{Current address: Department of Physics, University of North Texas, 76203 Denton, USA}
\author{Jorge I. Cerd\'{a}}
\affiliation{Instituto de Ciencia de Materiales de Madrid, ICMM-CSIC, Cantoblanco, 28049 Madrid, Spain}


\vspace{.15in}

\email[]{}

\date{\today}

\begin{abstract}
We use large-scale DFT calculations to investigate with unprecedented detail
the so-called spin-orbit (SO)
proximity effect in graphene adsorbed on the Pt(111) and Ni(111)/Au 
semi-infinite surfaces, previously studied via spin and angle resolved
photoemission (SP-ARPES) experiments. 
The key finding is that, due to the hybridization with the metal's bands,
the Dirac cones manifest an unexpectedly rich spin 
texture including out-of-plane and even radial in-plane spin components
at (anti-)crossings where local gap openings
and deviations from linearity take place.
Both the continuum character of the metallic bands and 
the back folding associated to the moir\'e patterns enhance the spin texture
and induce sizeable splittings which, nevertheless, only become giant 
($\sim$100~meV) at anti-crossing regions; that is, where electronic
transport is suppressed. At the quasi-linear regions the splitted bands
typically disperse with different broadenings and tend to cross
with their magnetization continuously changing in order to match that
at the edges of the upper and lower gaps. As a result, both the splittings and
spin direction become strongly $k$-depedendent.
The SO manifests in an analogous way
for the spin-polarized G/Au/Ni(111) system, although
here the magnetic exchange interactions dominate inducing small
splittings ($\sim$10~meV) in the $\pi$-bands while the SO mainly introduces
a small Rashba splitting in the Dirac cones as their magnetization
acquires a helical component. 
While revealing such complex spin texture seems challenging from the experimental side, 
our results provide an important reference for future SP-ARPES measurements of similar graphene
based systems extensively investigated for applications in spintronics.

\end{abstract}

\maketitle
\section{Introduction}   
Spin-orbit coupling (SOC) in graphene\cite{spintronics} has been proposed as a basis for various phenomena of fundamental and practical importance for spintronics,
such as the spin Hall effect,\cite{hydrogenation-exp,copper} 
topological quantum spin Hall effect (QSHE),\cite{kanemele} 
quantum anomalous Hall effect (QAHE),\cite{qah,qah-fe} 
weak localization\cite{weakloc} or electron confinement associated to
multiple topologically non-trivial gaps.\cite{calleja} Given the tiny intrinsic SOC of graphene (G), estimated to be less than few tens of $\mu$eV,\cite{fabian-dft, fabian-tightbinding} which makes the experimental realization of such phenomena unfeasible, extensive efforts have been devoted to find an efficient way to enhance and tune the strength of SOC extrinsically. One of the most promising
approaches is the so-called proximity effect\cite{proximity, fluorine, weeks, tmds}
whereby the large SOC of heavy elements either adsorbed or present in the
substrate may be transferred to the G, as can be explained in terms of diverse 
hopping processes onto and off the metal atoms so that the electron acquires a
large SOC before it returns to the graphene.~\cite{weeks} 
For any practical purposes, however, a straightforward and robust approach for SOC engineering is of vital importance. Epitaxial graphene grown on metallic surfaces seems to be an excellent candidate to achieve this goal, mainly due to an easy fabrication process and absence of a structural disorder which could 
deteriorate the spin and charge transport performance. Notably, metallic 
substrates also offer a chance to adsorb or intercalate atoms/layers providing 
a plethora of options for further tuning other relevant properties apart from 
solely changing the strength of SOC.\cite{calleja, kralj}
 
Despite continuous experimental attempts performed during the last years in 
order to achieve strong spin-orbit proximity effect in graphene,\cite{first13mev, main_nat, platinum, platinum2, copper, proximity, calleja, ir, hedgehog, misha} 
the corresponding theoretical studies are rather scarce and involve
oversimplified models mainly due to the large computational resources that
the inclusion of SOC requires. Typically, these works are restricted to either 
diluted metallic layers or slabs with small unit cells which cannot 
appropriately take into account the commensurability between graphene layer and
the underlying substrate. In fact, a realistic {\it ab initio} modeling of 
SOC-related phenomena in graphene/metal interfaces has been missing so far. In 
this paper we fill this gap by addressing, from a theoretical perspective, the 
fundamental question of how a semi-infinite metallic substrate alters the 
graphene's Dirac cones (DCs) and determines their spin texture. We consider 
graphene/metal interfaces involving large supercells
thus reducing artificial strains typically imposed when oversimplified 
commensurabilities are assumed. Therefore, our approach introduces
a further source of graphene/metal hybridization due to the Brillouin zone 
(BZ) backfolding whose impact, as shown below, is by no means negligible. 
We analyze two different examples of metallic substrates: a Pt(111) surface and
a Au monolayer adsorbed on a Ni(111) surface, the choice
being motivated by several spin- and angle-resolved photoemission spectroscopy (SP-ARPES) measurements where large Rashba splittings
(RS) of up to $\sim$100~meV have been reported for both systems\cite{first13mev, main_nat, platinum, platinum2} as
well as by the fact they present fundamental differences in their electronic
properties.
The former involves $\pi$-$d$ interactions between the graphene and the Pt
and, being non-magnetic, preserves the Kramer's degeneracy. It 
thus represents an ideal system to estimate the SOC derived splittings
by inspecting the spin structure of the Dirac cone states.
In contrast, the G-Au interactions are
mainly of the $\pi$-$sp$ type and, more interestingly, the gold layer shows a 
net magnetization due to the presence of the ferromagnetic substrate below which
constitutes an excellent playground to explore the interplay between SOC and 
spin-exchange derived splittings.\cite{rader} 
Furthermore, the same group has reported the co-existence of two phases for this 
system showing small and large splittings~\cite{main_nat}, although the origin of such 
puzzling difference could not be confirmed.
Throughout, we focus
on the induced spin-texture of the Dirac cones 
rather than on any topological properties that could emerge at
opened gaps (due to intrinsic SOC\cite{weeks,kanemele}) since we are
primarily interested on the spin properties of the linear $\pi$ bands.  
It turns out that the key mechanism behind the transfer of SOC to the
G's $\pi$-bands is hybridization with the surface localized metal $d$-states which
hold the largest SOC derived spin splittings. It is precisely at such (anti-)crossings
where mini-gaps are opened and the DCs show the largest distortions attaining band 
splittings above 100~meV. On the other hand, in the absence of strong hybridizations 
(regions of highly dispersive $s-p$ metal bands) the influence of the SOC on the G is 
minor and the splittings are at most a few tens of meV.

The paper is organized as follows: in Sec.II we discuss the details of the
DFT and Green's functions calculations. Sec. III is devoted to the origin and 
peculiarities of the spin textures induced in graphene's Dirac cones by a 
Pt(111) substrate.
In Sec. IV the interplay between SOC and magnetic order in G/Au/Ni(111) 
is examined, while in Sec.V we summarize the main conclusions and perspectives. 

\section{Methods}
Our density functional theory (DFT) based calculations were performed with the 
{\sc GREEN} code~\cite{green,loit}
employing an interface to the {\it ab initio} {\sc SIESTA} package~\cite{siesta}. The 
exchange-correlation (XC) interaction was treated under the generalized gradient
approximation (GGA) following the parametrization of Perdew, Burke, and 
Ernzerhof,\cite{pbe} including spin polarization in all cases where Ni atoms
were involved. Dispersion forces were taken into account via the 
semi-empirical scheme of Ortmann and Bechstedt.\cite{vdw_ortmann}
The fully-relativistic pseudopotential (FR-PP) formalism~\cite{soc}
was employed to account for the SOC.
Core electrons were replaced by norm-conserving pseudopotentials of the 
Troulliers-Martin type, with core corrections included for the metal atoms in order 
to better describe the XC and SOC terms.\cite{soc}
The atomic orbital (AO) basis set consisted of Double-Zeta Polarized (DZP) 
numerical orbitals strictly localized --we set the confinement energy to 
100~meV.  Real space three-center integrals were computed over 
3D-grids with a resolution of $\sim$0.07~\AA$^3$ --equivalent to 
500~Rydbergs mesh cut-off. 
Brillouin zone integration was performed
over $k$-supercells of around (18$\times$18) relative to the G-(1$\times$1) 
lattice.

All considered graphene/metal structures were first relaxed employing two-dimensional 
periodic slabs involving several metal layers and the graphene on top. In the
case of G/Pt(111) we considered six Pt layers thick slabs and two different 
moir\'e patterns, the so-called $(2\times2)$ and $(3\times3)$ phases,\cite{moires, moires2, martingago}
described in detail in Sec. III. On the other hand, and
based on previous studies on the Au/Ni(111) surface,\cite{interface} we modeled
the G/Au/Ni system assuming a ($9\times9$)/($8\times8$)/($9\times9$) 
commensurability between the G, Au and Ni lattices, respectively, with the Au
layer intercalated between the G and a four Ni layers thick slab. 
Two different phases were
considered: one, where the Ni surface is unreconstructed and a second
one, where the top Ni layer presents a large triangular 
reconstruction\cite{interface} --its precise geometrical description is 
given in the Appendix~C. In all cases
the graphene atoms and the first two metallic layers
were allowed to relax until forces were below 0.03~eV/\AA\ while the bottom 
layers (four in the case of Pt and two for Au/Ni) were held fixed to bulk 
positions. 

G adsorption energies were computed as the balance between the total
energy of the system and that of the sum of the clean relaxed metal surface and 
free-standing G. However, although the semi-empirical vdW approach followed here
is necessary to obtain the correct adsorption geometries, it 
largely overestimates adsorption energies~\cite{mercurio}. Hence, 
we have estimated these energies after removing all vdW contributions~\cite{copc}.

The electronic structure for the semi-infinite surfaces was computed following 
several steps. Once the structures described above were optimized, we added
four (one) bulk-like layers of Pt (Ni) at the bottom of the slabs and recomputed
their Hamiltonians self-consistently first neglecting and next including SOC 
(see Ref.~\onlinecite{soc} for full details of the implementation). 
In the last step, we used the appropriate Hamiltonian matrix elements to 
stack the graphene and first metallic layers on top of bulk Pt(111) and Ni(111)
semi-infinite blocks via Green's functions matching techniques following the 
prescription detailed elsewhere.\cite{ysi2,loit}

Since for a semi-infinite system the absence of translational symmetry along
the surface normal does not
allow to evaluate the band structure via the diagonalization of the Hamiltonian matrix, we
compute instead equivalent $k$-resolved density of states projected on
different surface atoms, PDOS($\vec{k},E$). For a particular layer $I$,
its DOS projection is calculated from the system Green's function blocks,
$G_{IJ}$, connecting $I$ with itself and its neighbor layers $J$, according to:
\begin{equation}\label{eq:s1}
 PDOS_I(\vec k,E)= \frac{-i}{\pi} \sum_{\sigma,J} 
    Tr\{ G^{\sigma\sigma}_{IJ}(\vec k,E) \; O_{JI}(\vec k)\}
\end{equation}
where $\sigma=\uparrow,\downarrow$ denotes the spin
component and $O_{IJ}(\vec k)$ stands for the $k$-space overlap between layers 
$I$ and $J$. The summation over $J$ only includes layers $I-1$, $I$ and $I+1$ 
because all layers are defined thick enough so that overlap matrices beyond 
first nearest neighbor layers vanish (obviously, for the surfacemost layer
only the $I$ and $I-1$ terms will enter the above equation).

Similarly, $k$-resolved magnetization densities,
$\vec m(\vec k, E)$, may be obtained as:

\begin{eqnarray}
 m_x(\vec k, E)= \frac{-2 i}{\pi} \sum_J
    Tr\{ G^{\uparrow\downarrow}_{IJ}(\vec k, E)
           \; O_{JI}(\vec k)\}  \\
 m_y(\vec k, E)= \frac{2}{\pi} \sum_J
    Tr\{ G^{\uparrow\downarrow}_{IJ}(\vec k, E)
           \; O_{JI}(\vec k)\}  \\
 m_z(\vec k, E) = \\ \nonumber
\frac{-i}{\pi} \sum_J
    Tr\{ \left( G^{\uparrow\uparrow}_{IJ}(\vec k, E)-
          G^{\downarrow\downarrow}_{IJ}(\vec k, E) \right)
           \; O_{JI}(\vec k)\}
\end{eqnarray}

In this work, we will present most of our results in the form of \ke\ maps
projected either on the G or the metal layers. Furthermore, most of the G
and first Pt layer projections have been computed within the moir\'e supercell 
(folded electronic/magnetic structures) as well as assuming that 
translational (1$\times$1) symmetry is preserved within the layer (unfolded
structures) following the approach described in the Appendix~A.
For all \ke maps we have typically
employed a resolution of $\sim 6\times10^{-3}$~\AA$^{-1}$ in $k$-space and
1-2~meV in energy while the imaginary part of the energy entering the Green's
function calculation (self-energy or broadening) was accordingly set to 2-4~meV.
These small values ensure that the widths of the peaks in the calculated
G PDOS arise from the metal's self-energy (i.e. the interaction with the
continuum of metal bands). We note that such a high
resolution required a huge computational effort, as the maps presented in 
this work typically comprised of the order of $10^6$ and $10^5$ \ke grid points
for the G/Pt and G/Au/Ni systems, respectively.

\section{Graphene on Pt(111)} 
\begin{figure}
\includegraphics[width=0.45\textwidth]{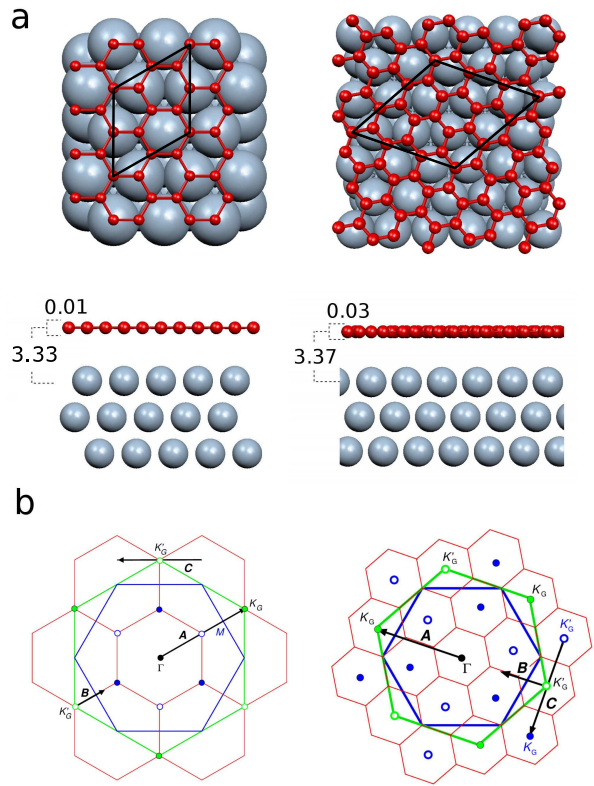}
\caption{\label{geom}
(a) Top and side views of the G-(2$\times$2)/\retwo\ (left-hand panel) and G-(3$\times$3)/Pt(111)-($\sqrt{7}\times\sqrt{7}$)$R19.1^{\circ}$ (right-hand panel) configurations.  
C and Pt atoms are represented by small red and large blue
balls, respectively. The commensurate supercells are indicated by the 
parallelograms. In the side views the optimized Pt-G interlayer distance and
the G's intralayer corrugation are given in \AA.
(b) Corresponding combined BZ schemes for each phase. Small red hexagons 
correspond to the G's folded BZ --(2$\times$2) at the left and (3$\times$3) at
the right, while larger blue and green hexagons correspond to the Pt(111) and 
G's primitive BZs, respectively. Closed/open blue circles mark the G's
$K_G$/$K'_G$ points including those backfolded into the Pt's primitive BZ.
Black arrows indicate the $k$-lines $A$, $B$ and $C$.}
\end{figure} 

\begin{figure*}
\includegraphics[width=\textwidth]{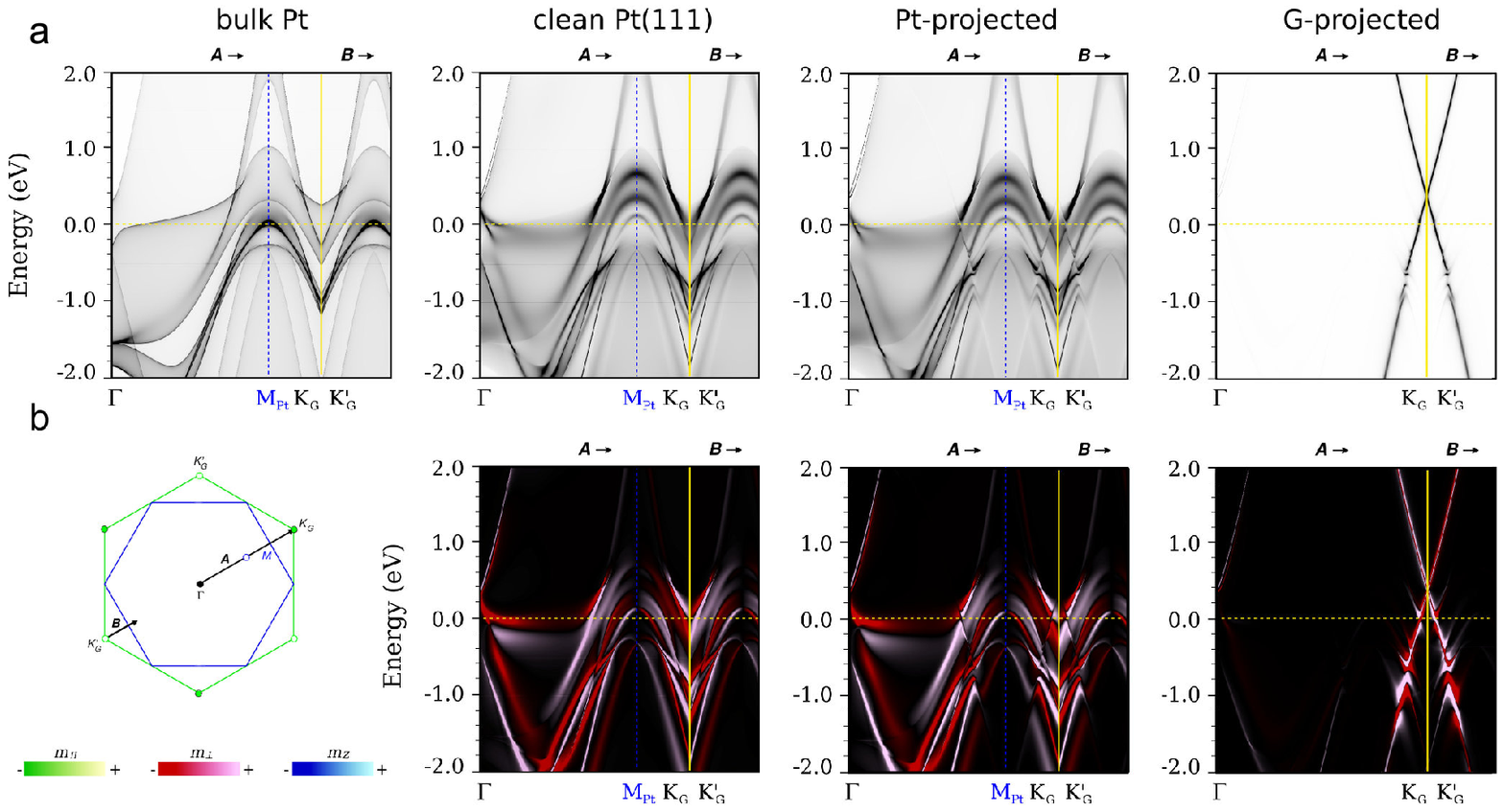}
\caption{\label{2x2-gk}
(a) First column: PDOS\ke map of bulk Pt along the A-B $k$-path defined in the 
inset below and Fig.~\ref{geom}(b). Second column: Same as first column, but 
calculated for clean Pt(111)-(1$\times$1) semi-infinite surface and projected on
the surface layer. Third column: PDOS\ke map calculated for the 
G-(2$\times$2)/\retwo\ semi-infinite surface and projected on the Pt surface
layer along the same A-B $k$-path after unfolding onto the Pt (1$\times$1) BZ. 
Last column: Same as third column but projected on the graphene layer after
unfolding onto the G's (1$\times$1) BZ. 
(b) The associated G's magnetization map, $\vec m$\ke, for the same systems
and projections as in (a) after superimposing the $\perp$, $\parallel$ and 
$z$ components, each color coded as indicated by the legends at the bottom left
(although only \mper\ is non-zero along these $k$-paths). 
The inset on the left shows the primitive BZs of graphene (green) and Pt(111) 
(blue) and the considered high-symmetry $k$-paths.  
}
\end{figure*}
\begin{figure}[ht]
\includegraphics[width=\columnwidth]{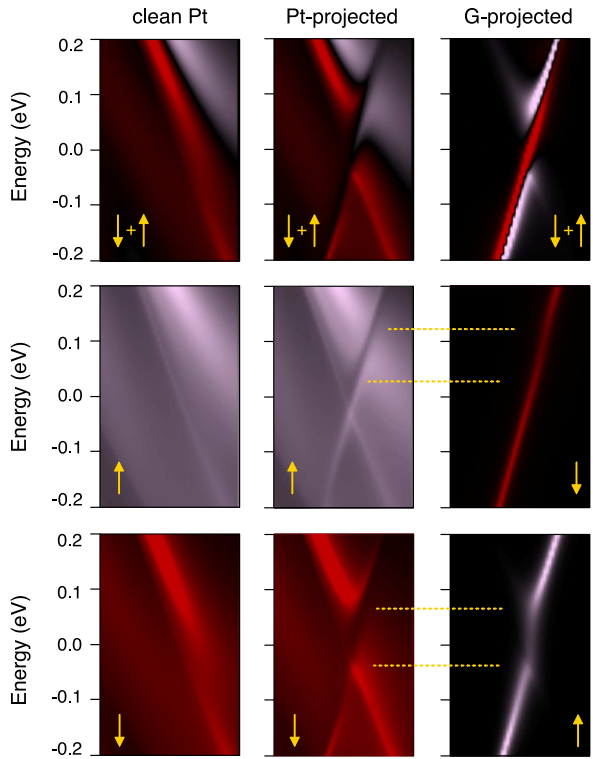}
\caption{\label{zoom1}
Top row: zoom in of the magnetization maps shown in Fig.~\ref{2x2-gk}(b)
in the vicinity of $E_F$. First
column corresponds to the clean Pt(111) surface, and second and third columns
to the G and Pt projections, respectively. Middle and bottom rows present
a decomposition of the \mke\ maps into its \up\ (pink) and \down\ 
(red) components --see arrows in the inset of each figure. 
Since the G-Pt coupling in this region is anti-ferromagnetic,
we have placed in each row the G's spin component opposite to that of the Pt.
Yellow dashed lines indicate the lower and upper edges of the gaps opened at
each spin branch.}
\end{figure}
\begin{figure}[ht]
\includegraphics[width=\columnwidth]{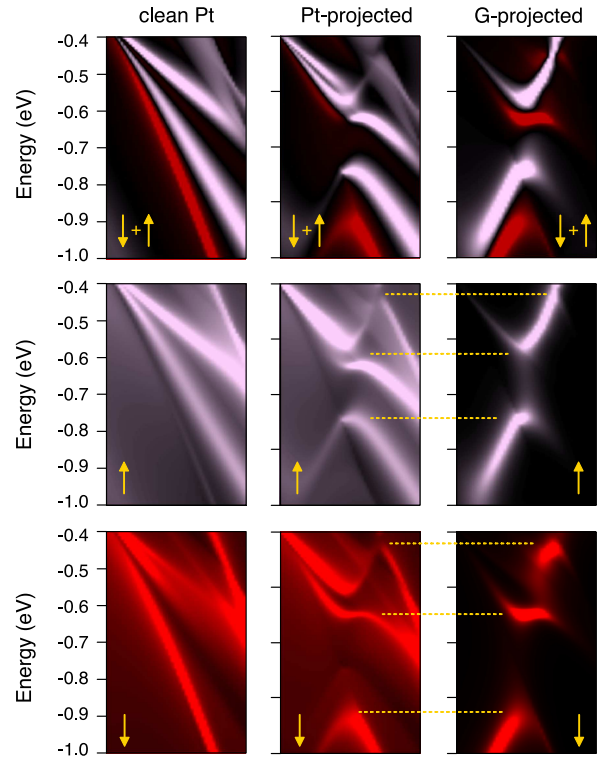}
\caption{\label{zoom2}
Similar zooms as in Fig.\ref{zoom1} but taken over a wider energy range below
$E_F$. This time, since the Pt-G coupling is mainly ferromagnetic we have
placed the \up\ or \down\ projections for both G and Pt in the same rows.
}
\end{figure}
\begin{figure*}[ht]
\includegraphics[width=\textwidth]{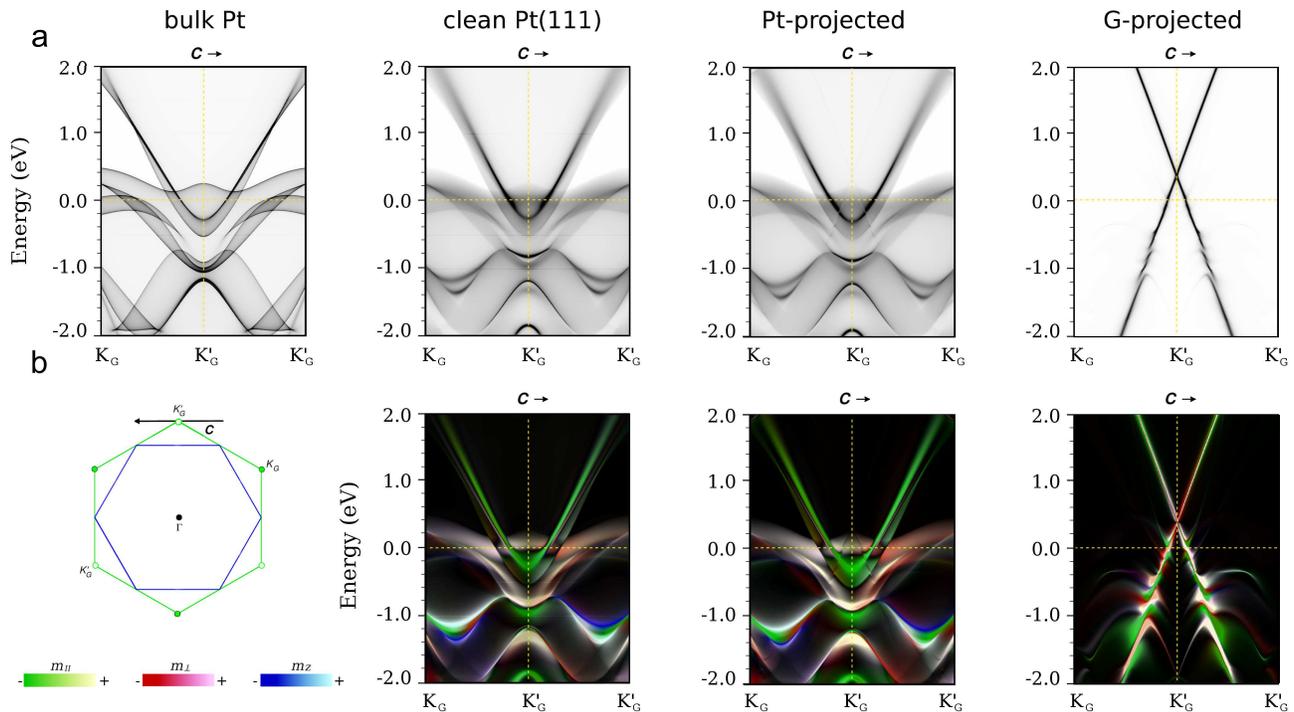}
\caption{\label{2x2-ppd}
Same as Fig.~\ref{2x2-gk} but calculated along the direction perpendicular to 
$\Gamma-K'_{G}$ ($k$-line $C$ in the inset).
To view each spin component independently see Fig. S1 of the Supplementary Material.
}
\end{figure*}

\begin{figure*}[ht]
\includegraphics[width=\textwidth]{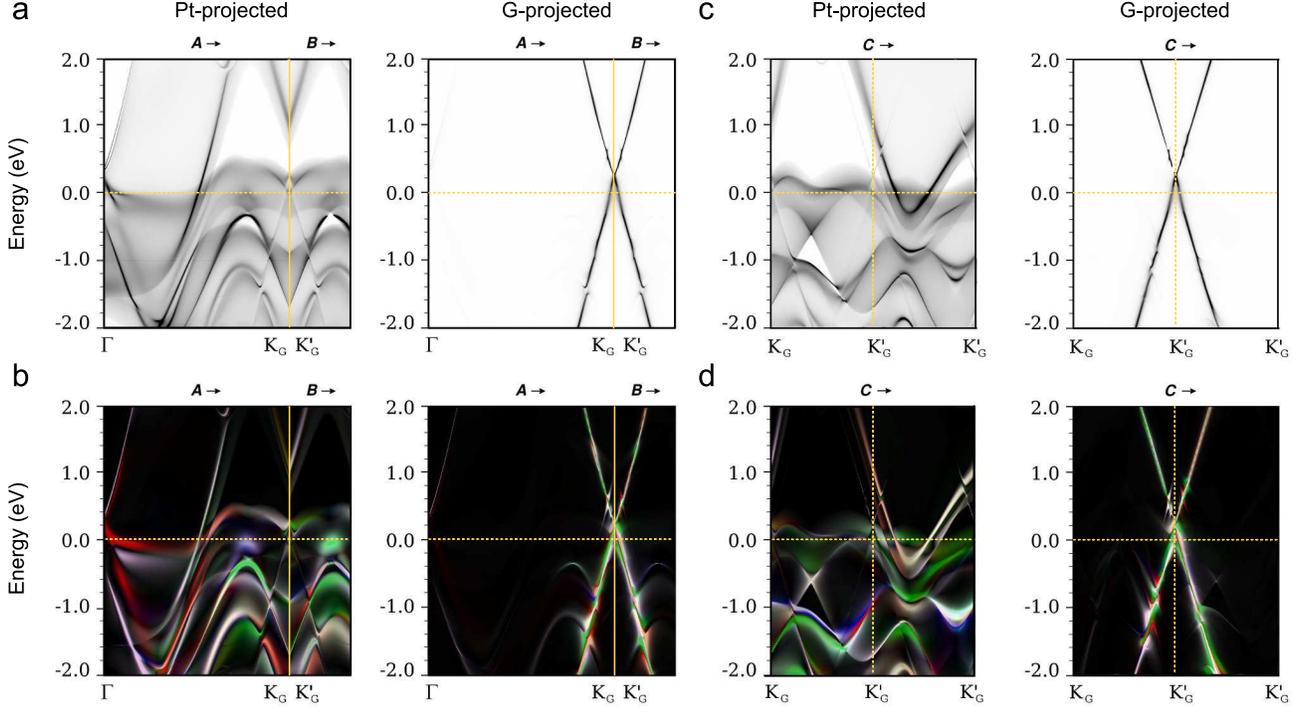}
\caption{\label{3x3}
(a) Left-hand panel: PDOS\ke map projected on the Pt surface layer calculated 
for the G-(3$\times$3)/\rethree\ semi-infinite surface and unfolded onto the
Pt (1$\times$1) BZ along the $k$-lines $A$ and $B$ indicated in 
Fig.~\ref{geom}(b). Right-hand panel: Same as (a) but projected and unfolded on
the G's primitive BZ. 
(b) Corresponding magnetization density maps following the same 
color scheme same as in Fig.~\ref{2x2-ppd}; this time all three components of 
\mke\ are non-trivial. 
(c) Same as (a), but calculated along path $C$ defined in
Fig.\ref{geom} (b). (d) Spin texture corresponding to the PDOS 
shown in (c).
To view each spin component independently see Figs. S2 and S3 of the Supplementary Material.
}
\end{figure*}

\begin{figure*}[ht]
\includegraphics[width=\textwidth]{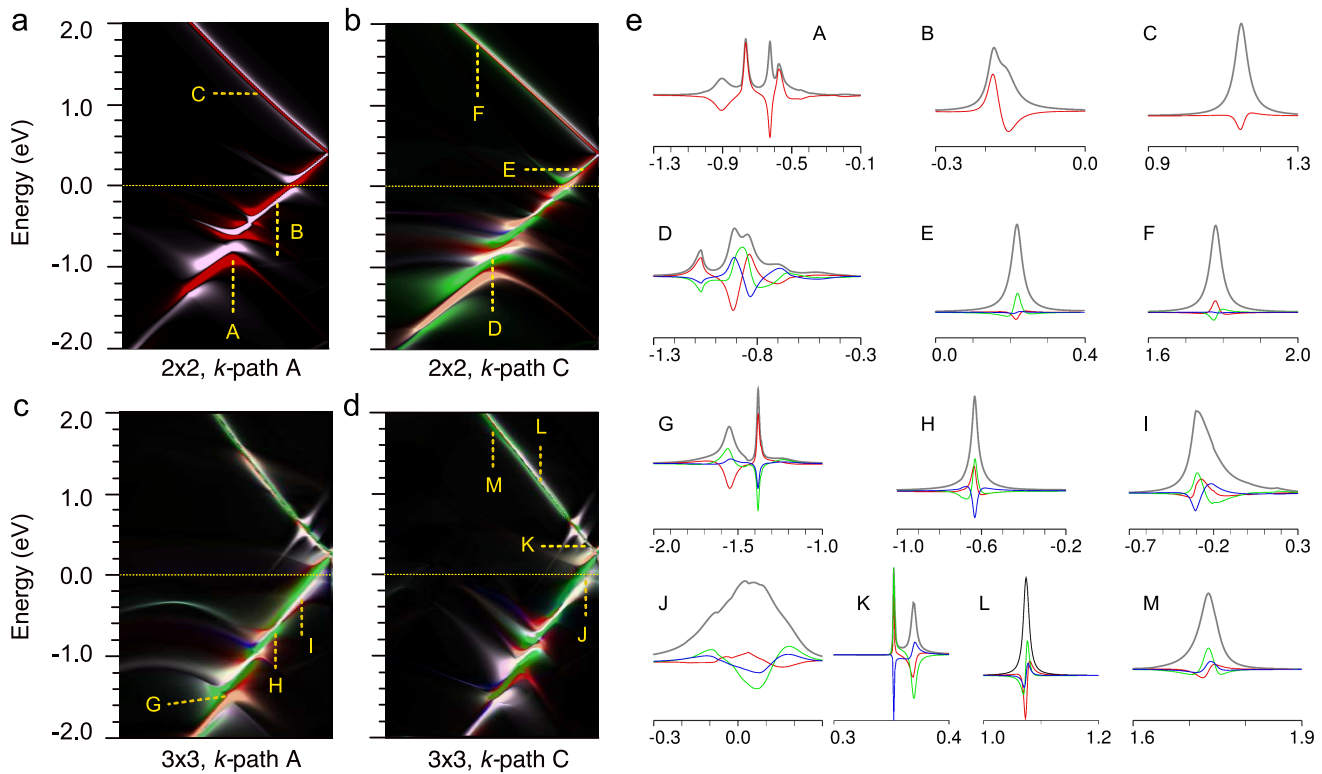}
\caption{\label{peaks}
(a) Zoom in of the G's projected magnetization map in the (2$\times$2) phase
presented in last column of 
Fig.~\ref{2x2-gk} covering part of the A $k$-path. (b) Same as (a) but for the 
\mke\ map of Fig.\ref{2x2-ppd} ($k$-path C). (c)-(d) Same as (a)-(b) but
for the G-magnetization maps of Fig.\ref{3x3} corresponding to the (3$\times$3)
phase. (e) PDOS($E$) and $\vec{m}(E)$ single spectra extracted from these
maps at specific $k$-points marked with capital letters (A to M) in panels 
(a-d). Grey, red, green and blue lines represent the PDOS, and $\perp$, 
$\parallel$ and $z$ components of $\vec{m}$, respectively. Note that the energy
ranges considered in the plots can be rather different among them, ranging from
1~eV down to 100 meV while the $y$-axis (not shown) has been re-scaled in each 
plot for the sake of clarity (indeed, the intensity of the sharp peaks at the 
upper DC tends to be a factor 3-5 larger than in the lower cone).}
\end{figure*}

Out of the over 20 
different moir\'e patterns reported for this system,\cite{moires, martingago} 
we considered the two most common phases namely, 
(2$\times$2)/\retwo\, and (3$\times$3)/\rethree --hereafter denoted as
(2$\times$2) and (3$\times$3), respectively. As described above, both were modeled
placing a graphene sheet on top of a Pt(111) slab and fixing the Pt bulk lattice constant
to its experimental value, $a_{Pt}=3.92$~\AA, leading to G's lattice constants
of $a_G=2.40$ and 2.44~\AA\  for the (2$\times$2) and (3$\times$3) supercells,
respectively. Whereas the former represents a noticeable 2.5~\% 
compression with respect to pristine G, the latter is only 0.8~\%
smaller than the experimental value of 2.46~\AA.
In Fig.~\ref{geom}(a) we show the optimized geometries, for which we obtained
an uncorrugated G layer adsorbed at 3.33 and 3.36~\AA\ from the Pt surface, in
good agreement with the 3.3~\AA\ distance obtained 
experimentally.\cite{sadowski} Energetically, the ($3\times3$) phase was
found marginally more stable than the ($2\times2$), with adsorption energies (see
section Methods) of 26
and 23~meV/C, respectively, both clearly in the physisorption regime.\cite{gold-rpa}

Figure~\ref{geom}(b) sketches the 2D reciprocal space for both phases with
the G and Pt BZs indicated in green and blue, respectively, and that of the
commensurate supercell in red in an extended zone scheme. 
Whereas in the (2$\times$2) case the $K_G$ and $K'_G$ high-symmetry points 
are back-folded into the supercell's BZ $K'$ and $K$ points, for the 
(3$\times$3) they all transform into the $\Gamma$ point. Nevertheless,
the quasi-freestanding character of the graphene layer allows us to accurately unfold the PDOS\ke 
and \mke\ quantities onto its primitive BZ and, hence, examine the $\pi$-band 
dispersion and spin texture around $K_G$ and $K'_G$ separately 
(see Appendix~A).
Similarly, the scarce reconstruction of the Pt topmost layers permits 
an analogous unfolding but onto the Pt(111)-(1x1) BZ.
We have computed the electronic structure for $k$-lines running along the $\Gamma-K_G/K'_G$
direction (paths $A$ and $B$ in Fig.~\ref{geom}(b)) and a third line passing 
through $K'_G$ but perpendicular to the previous ones (path $C$).

Since the (2$\times$2) phase is the most simple and symmetric one, we will
first present a very detailed study for this phase in order to establish the
main mechanisms dictating the G's induced spin texture; as shown in the next
subsections, most of them still hold for the ($3\times$3) case.

\subsection{(2$\times$2) phase}
Figure.~\ref{2x2-gk}(a) presents the unfolded electronic dispersion PDOS\ke 
for the (2$\times$2) phase projected on the first two Pt layers and on graphene along
the $A$ and $B$ paths (two rightmost columns in the figure). Both paths are displayed side by side because 
they are related through (2D) inversion symmetry thus yielding 
symmetric PDOS about the degenerate $K_G/K'_G$ point.
For the sake of clarity and completeness, we also show in the two leftmost
panels the electronic structure of bulk Pt and the clean 
Pt(111)-(1$\times$1) surface along the same $k$ lines.
The bulk projection shows the expected continuum of $d$-bands mainly
covering the occupied states region as well as a prominent bump at the Pt's BZ
boundary (point $M_{Pt}$) raising up to 1~eV. Narrow gap areas appear
across $\Gamma-M_{Pt}$ as well as around $K_G$.
The rest of the energy region
above the the Fermi level, $E_F$, is filled by a continuum of highly
dispersive (less intense) $sp$-type bands with large gaps emerging from
$\Gamma$ and $K_G$. At the edges the DOS
becomes sharp and intense due to the projection of the bands' curvature
along the surface normal. 
The PDOS\ke for the clean surface presents several important
differences with respect to its bulk counterpart: a smearing and broadening of 
the band edges, the appearance of intense surface resonances corresponding to 
states with a strong 2D character and the filling of all bulk gaps
below $E_F$. Moreover,
a sharp Rashba splitted surface state emerges from $\Gamma$ with its
onset just below 0.4~eV. 

The equivalent surface Pt projections computed for the
combined G/Pt(111) surface system are essentially the same as for the clean
case, except for 
quasi-linear traces belonging to the $\pi$-bands of graphene which can be seen around $
K_G/K'_G$ specially in the $d$-band region
(together with a back-folded replica at the left of $M_{Pt}$).
The PDOS\ke projected on graphene reveals that the Dirac cones are
essentially preserved, with a strong 0.40~eV $p$-type doping as
a consequence of the compressed C-C distances,\cite{gold-dft} while no gap
is opened between the upper and lower cones signaling a weak intrinsic SOC.
Within the 2~meV energy resolution (broadening) employed, no hints for any RS 
can be seen in the maps but just an overall blurring of 
the lower cone due to hybridization with the Pt-$d$ bands, as well as a large 
gap opening at $\sim-0.7$~eV. The upper cone, on the other hand, appears 
sharper and closely resembles that of pristine graphene.

Inspection of the spin structure enhances all these features and allows to
examine the SOC induced effects in greater detail. In Fig.~\ref{2x2-gk}(b) we 
present the magnetization dispersion \mke\ for the same projections as in (a).
The magnetization vector has been decomposed into two inplane components,
one along the $k$-line (\mpar) and the other one orthogonal to it (\mper)
plus the the out-of-plane contribution ($m_z$). 
The \mke\ map for the bulk phase is omitted since Kramer's degeneracy 
($E(\vec k,\sigma)=E(-\vec k,\sigma')$) when combined with the
inversion symmetry holding for $fcc$ Pt ($E(\vec k,\sigma)=E(-\vec k,\sigma)$) 
forbids any net magnetization in $k$-space --indeed, the computed \mke\ map is
completely dark. On the other hand, and as
expected for a heavy metal surface,\cite{mazzarello}
the clean Pt(111)-(1$\times$1) spin texture is very rich presenting large
splittings associated to magnetizations which are always locked along the \mper 
direction (only red-pink tones appear in the maps). The spin locking results from
the fact that
the $A$ and $B$ paths run along a mirror symmetry plane so that the symmetry 
transformation applying to the $\vec m$ {\it pseudo-vector}, 
$\tau(m_\parallel,m_\perp,m_z)=(-m_\parallel,m_\perp,-m_z)$, leads to vanishing
$m_\parallel$ and $m_z$ components precisely at the mirror plane.
On the other hand, due to Kramer's degeneracy, the Rashba splitted bands
are anti-symmetric about the time reversal invariant momentum (TRIM)
$M_{Pt}$ point, with abrupt inversions of \mper\ occurring
at this point. Likewise, the same
antisymmetric behaviour is found between the $A$ and $B$ paths
around the $K_G/K'_G$ points. 

The Pt magnetization dispersion hardly changes 
when the graphene is adsorbed on top, although the traces of Dirac states become
more patent and can be identified as mini-gap openings at the avoided crossings
with the Pt's Rashba splitted bands (see below). Notably, the G-projection
shows an unexpectedly complex spin texture, with the lower cone 
covered with streaks and undergoing numerous spin flips due to 
hybridization with the Pt bands while the upper cone appears free of any 
crossings. Nevertheless, both the Pt and G magnetizations
remain locked to the momentum (only \mper\ component) since the mirror
symmetry plane is preserved in the combined G/Pt (2$\times$2) phase.

In order to gain insight into the SOC mediated graphene/Pt hybridization 
mechanism we first examine the energy region where the metallic bands
cross the $\pi$-bands closest to the Dirac point (0.3~eV interval around $E_F$).
A zoom in of this region is presented in 
Fig. \ref{zoom1}, including the Pt and G \mke\ projections (second and 
third columns, respectively) as well as that for the clean Pt surface (first 
column). We have further decomposed the magnetization into its \up\ and \down\ 
components with their respective maps appearing in lower rows
(to this end, we diagonalized the (2$\times$2) PDOS$^{\sigma\sigma'}$ matrix at
each energy and $k$-point).
The clean Pt electronic structure in this region consists of a broad \up\
and an intense and sharper \down\ continuum of bands, the former slightly 
shifted to the right. In the combined G/Pt system each spin component of 
graphene interacts with the opposite spin component metal band tearing it after
opening a gap across it (marked by the dashed yellow lines). Therefore, the spin
coupling in this region is antiferromagnetic.
Each $\pi$-branch itself also opens a $\sim$100~meV gap as
revealed by clear deviations from linearity at the upper and lower 
gap edges, especially at the G's \up\ branch, which interacts more strongly. 
Indeed, the gap edges can be clearly resolved in the corresponding Pt 
projections while a blurred DOS crosses the gap for both 
graphene's spin components, although at this point we cannot discern if it 
corresponds to a topologically protected state or to slowly decaying evanescent
waves or to some type of {\it partial} hybridization.
Due to the Rashba shift between the \up\ and \down\ Pt bands, the location of 
the gap differs in energy between both spin components by $\sim$100~meV.
The resulting G's magnetization acquires a complex texture;
above the gap both spin components run in parallel albeit the \up\ branch 
is broader and slightly $k$-shifted to the left. Below the avoided crossing, 
both emerge with similar broadenings but now the \up\ $k$-shifted to 
the right of the \down\ branch.

Figure~\ref{zoom2} displays another zoom in of Fig.~\ref{2x2-gk}(b) 
covering the -1.0 to -0.4~eV energy interval where the $\pi$-bands undergo the 
largest distortions. Here the G's \up\ branch interacts 
with the Pt's leftmost \up\ band (ferromagnetic coupling) opening a large
gap of almost 200~meV which is again crossed by a blurred DOS.
The G's \down\ branch is however much more textured, as it opens
three gaps with vanishing DOS across them. The lower gap edge (around $-0.7$-eV)
arises from the ferromagnetic coupling with the sharp leftmost Pt \down\ branch.
Slightly below $-0.6$~eV a saddle-point feature develops (interband state) which
links the \down\ Pt branch at its left with another \up\ band at its right, 
leaving an abrupt inversion of the magnetization which can be clearly seen in 
the Pt-projected map at the top row. Another intense G \down\ feature appears 
at $-0.45$~eV as result of the hybridization with the Pt band dispersing at
the upper right corner. Since this band is hardly spin polarized (it contains
similar \up\ and \down\ contributions) it also hybridizes with the G's \up\ 
component creating a minigap which can be more clearly seen in the \up\ 
Pt-projected map.

Once we have shown for the simplest case --spin locked in one
direction-- that the transfer of SOC from the metal to the G $\pi$-bands
occurs at anti-crossing regions and is strongly spin dependent, we next 
generalize this conclusion to more complex cases by gradually lifting the 
symmetry constraints. Figure~\ref{2x2-ppd} displays 2D maps analogous to 
those in Fig.~\ref{2x2-gk}, but computed along $k$-path $C$. Since this line
is perpendicular to the mirror plane, the magnetization is not
constrained along the $\perp$ direction any more and its three components are in
general non-zero. They have been simultaneously merged in each frame employing 
green, red and blue color scales for \mper, \mpar\ and $m_z$, respectively (see
color scheme in the legend of Fig.~\ref{2x2-ppd}(b)).
The last two are antisymmetric about $K'_G$ with abrupt inversions 
occuring at this point, while the DOS\ke and \mper\ component
remain symmetric. In the Pt projections
an intense parabolic band already present in the bulk (onset from $K'_G$ at 
$\sim -0.3$~eV) disperses across the empty states region, while 
surface resonances appear at the $d$-band edge crossing $E_F$ and
sharp surface $d$-states develop around -1.0~eV where the bulk presents
gap areas. They all become more patent in the magnetization maps, which
show that the spin is mostly confined in-plane ($m_z$ tends to be smaller)
except at the $d$-band minima located at -1~eV, which acquire a strong
out-of-plane character. 

Similar to the $A-B$ path, the upper Dirac cone
remains almost intact (absence of hybridization with any metallic bands), sharp
and hardly splitted within the high 2~meV energy resolution (broadening) employed in our calculations. 
Its spin remains in-plane ($m_z=0$) but, at contrast to
the previous $A-B$ path, it is not
strictly tangential to the cone since the \mpar\ projections do not 
vanish.
The lower cone, on the other hand, is strongly distorted due to
multiple anti-crossings with the Pt bands. In the PDOS map we can notice two
large gaps which open at approximately -0.7 and -1.1~eV. Furthermore, despite the complexity of the
color scheme employed to visualize $\vec m$, spin re-orientations 
can be readily identified throughout.

One must recall, however, that not all the induced streaks and gaps in the DCs
can be ascribed to anticrossings with the unfolded metal bands shown in 
Fig.~\ref{2x2-ppd}. In the Appendix~B we present side by side folded and unfolded
G+Pt projections obtained along the same $k$-line (Figure~\ref{fold}). 
Due to the BZ back-folding
associated to the (2$\times$2) supercell additional Pt bands in other regions
of the Pt(111) (1$\times$1) BZ appear in the folded map that, indeed, cross
the $\pi$-bands --also recall the replica of the DC traces visible away from
$K_G$ in the Pt-projected maps of Fig.~\ref{2x2-gk}.

\subsection{(3$\times$3) phase}
 Fig.~\ref{3x3} summarizes analogous maps calculated for the (3$\times$3)
configuration along the paths marked in Fig. \ref{geom}(b). 
Since this phase does not present any mirror planes (it
belongs to the $p3$ symmetry group), there are no restrictions to the
magnetization components. Kramer's degeneracy leads to symmetric 
PDOS and antisymmetric $\vec m$ maps between the $K_G-\Gamma$ and $K'_G-\Gamma$
directions (paths $A$ and $B$ in Fig.~\ref{3x3}(b)), while along path $C$
the structure results highly asymmetric. Otherwise, the overall SOC mediated 
interaction picture is very similar 
to the previous phase illustrated in Fig.~\ref{2x2-ppd}. 
Intense surface states/resonances with well defined magnetizations and
oriented along multiple directions (as can be inferred from the highly
polycrhomatic maps) decorate the BZ up to $\sim 0.5$~eV above $E_F$.
Their (anti)-crossings with the Dirac cones leads to a blurring of
the PDOS and the opening of multiple gaps (notice a particularily large one at
$-1.5$~eV in paths A and B) which transfer a highly complex
spin texture to the G states. The upper DC is sharper since only a few
Rashba splitted metal $sp$-bands disperse across the empty states
region but, still, its spin texture is non trivial.
A sensible
difference with the (2$\times$2) case is the larger number of back-folded
metal bands that interact with the DCs as a consequence of the larger size
of the moir\'e pattern; indeed, two bands cross this time the previoulsy
unperturbed upper cone
close to the Dirac point (DP). This fact is highlighted in the folded versus unfolded
comparison shown in the Appendix~B (Figure~\ref{fold}).

\subsection{SOC induced splittings in the G-bands}
We complete the G/Pt analysis by presenting in Figure~\ref{peaks}
zooms in of the G projected magnetization maps for both phases and along 
$k$-paths A and C (left panels (a-d)). Single spectra extracted from these
maps at some representative $k$-points are shown in panel (e) (colored lines), 
including as well the corresponding PDOS$(E)$ curve (grey lines). Let us
first consider the anti-crossing regions; plots A, D or G. Here,
giant splittings larger than 100 meV can be readily identified in most of
the PDOS. The curves comprise up to four peaks, two belonging to the
lower edge of the gap and the other two to the upper one. However, since
the broadening of each peak can be quite different due to a stronger 
hybridization of one of the spin components with the metal bands, they are 
often hard to resolve in the spectra --specially in G. In general, and if not
forbidden by symmetry, the three components of the magnetization are non-zero 
with the spin orientation of each graphene branch
depending on that of the metal band causing the anti-crossing. 
It is particularly suprising the emergence of large \mpar\ 
components (that is, in-plane radial contribution) indicating that no
spin-momentum locking at the DCs holds.

To understand the spin rotation process and how the size of the splittings 
changes between the gap regions let us focus on the sequence of spectra H-I; 
at H there is a sharp state with $\vec m=(+m_\perp,+m_\parallel,-m_z)$ and a 
broader one with the opposite orientation. Since at this particular $k$-point
both are aligned, there is no net splitting.
As we move upwards in energy along the lower cone the fomer broadens
and shifts to the left leading to a sizeable splitting of $\sim$100~meV --as may
be deduced from the distance between the $m_z$ mimimum and maximum. 
It is also clear that the magnetization changes direction as the band disperses.
Other curves acquired at similar quasi-linear regions in the lower cone
follow a similar pattern (B or E). The fact
that their corresponding PDOS show a single (or at most an asymmetric) peak 
rather than two splitted ones is reminiscent of the resolution achieved in 
ARPES versus SP-ARPES experiments, since in the former case, typically, no 
splittings can be resolved in the $\pi$-bands. 

Spectra J and K, taken at the same $k$-point close to the Dirac point, 
show the drastic transition between the lower and upper
cones whereby a single broad and complex peak in the PDOS appears
as two sharp maxima splitted by a few tens of meV above $E_F$. Indeed, in
the  empty states regions there are hardly any anti-crossings and the bands
remain sharp and fairly linear (curves C, F, L and M) while the magnetization 
does not undergo such abrupt changes (albeit the \mpar\ component is still
comparable to the other two). Nevertheless, their splittings become of 
the order of just 10~meV.

We complete our discussion by comparing the calculated G/Pt properties
with the previous ARPES and SP-ARPES results reported by Shikin {\it et al.} and
Klimovskikh {\it et al.} in Refs. [\onlinecite{platinum,platinum2}] for
the (2$\times$2) phase as confirmed by the corresponding LEED patterns.
Overall, the experiments seem to agree quite well with our maps, specially
concerning the hardly perturbed Dirac cones after comparing their spectra
against our calculated G's PDOS in Fig.~\ref{2x2-gk}(a). Furthermore, 
the large calculated gap at $-0.7$~eV matches, within a DFT error of a few
hundreds of meV, a pronounced anticrossing between the $\pi$-bands
and the Pt $5d$-states appearing at around $-0.5$~eV (Figs.~1a and 4a in 
Refs.~[\onlinecite{platinum,platinum2}]). We also note a discrepancy between
the measured and calculated $p$-doping level; in the above mentioned
works the DP is shifted by 100 meV from $E_F$ while in our calculations it is 
close to 0.4~eV which is, however, only slightly larger than the $0.3-0.4$~eV
obtained in previous ARPES and STS experiments\cite{sadowski, ugeda} as well
as in other DFT studies under different XC functionals.\cite{giovanetti,fingerprints}.
Finally, in the light of the magnetization spectra shown in Fig.~\ref{peaks}(e),
we ascribe the giant splittings ($80-200$~meV) reported in these experimental 
works to or close to gap regions where such large values commonly appear. We
emphasize, however, that they cannot be interpreted as a standard Rashba-shift
betwen the \up\ and \down\ branches of the DCs whereby both spin components
would disperse in parallel around $E_F$. At contrast, the picture that emerges
from our simulations is that both the magnetization direction and the splittings
undergo continuous changes in $k$-space, while in the linear regions (upper
DC), the splittings are not larger than a few tens of meV.

\section{Graphene on Au/Ni(111)}
\begin{figure*}[ht]
\includegraphics[width=0.9\textwidth]{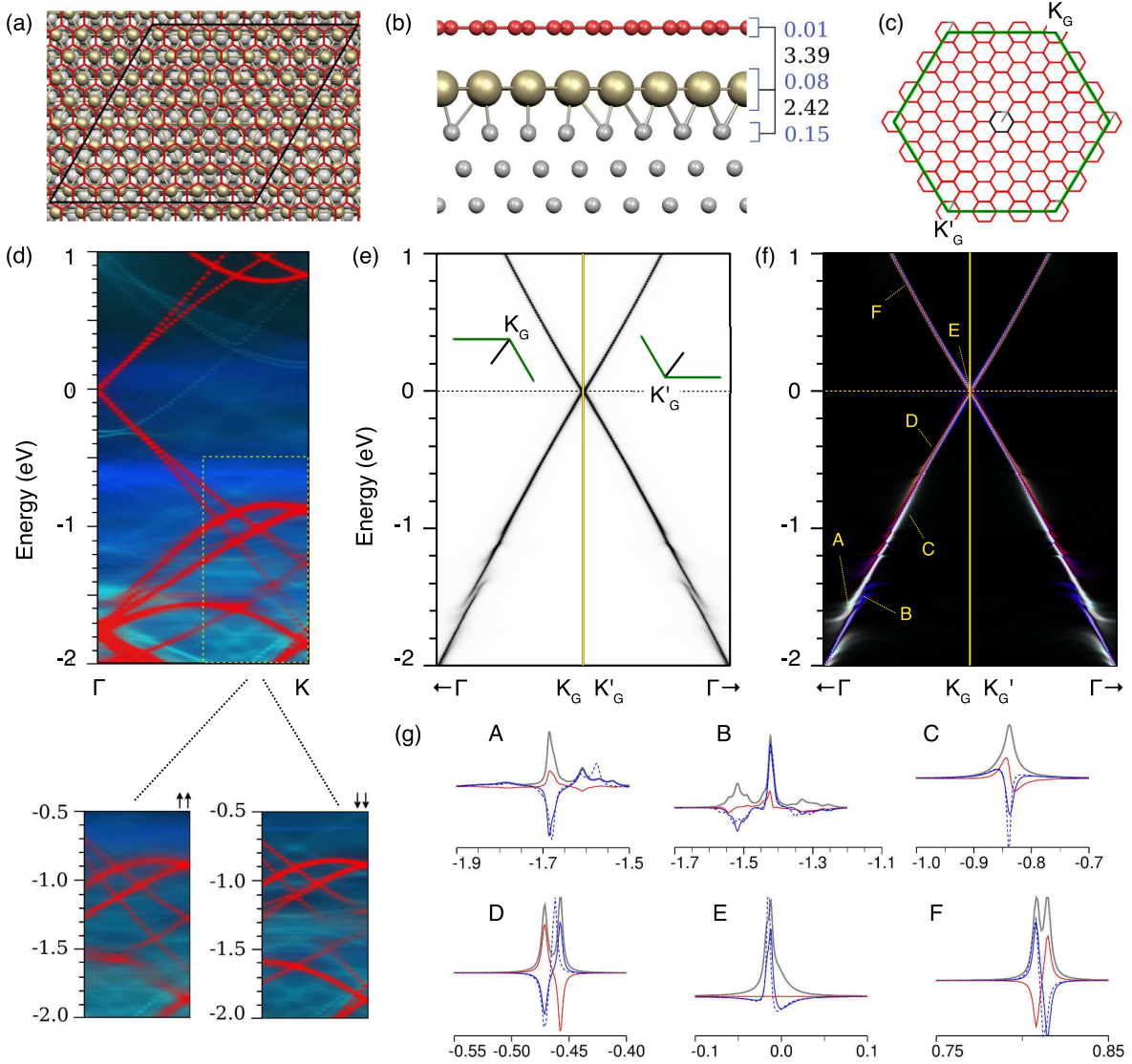}
\caption{\label{gold} Top (a) and side (b) views of the G/Au/Ni(111)
interface with C, Au and Ni atoms drawn in red, gold and grey, 
respectively. The parallelogram denotes the (9$\times$9)/(8$\times$8)/(9$\times$9) 
supercell employed. Interlayer average spacings and intra-layer corrugations
(blue font) are indicated in \AA. 
(c) BZ scheme of the system; black and red small hexagons indicate the 
supercell's BZ, while the large green hexagon corresponds to the G's 
(1$\times$1) BZ.
Labels with the subindex $G$ refer to the G's primitive BZ. 
(d) Folded PDOS\ke map calculated along the $\Gamma-K-M$ $k$-path indicated by 
a small solid line in (c) and simultaneously projected on the topmost surface 
layers of the semi-infinite system: graphene (red), 
Au (light blue) and the Ni surface (dark blue). The inset below shows the PDOS
decomposition into majority and minority spin components
($\uparrow\uparrow$ and $\downarrow\downarrow$ stencils) in the low energy
region. 
(e) PDOS\ke projected and unfolded on graphene along a fragment of the
$\Gamma-K$ path of the primitive G's BZ. 
(f) The corresponding magnetization density following the same color scheme
as in Fig.~\ref{2x2-ppd}; only \mper\ and $m_z$ components are present in this 
case. (g) Single spectra for the PDOS(E) (grey line) and $m_\perp/m_{z}$ 
components (red/blue) extracted at selected $k$-points marked in (f). 
Magnetization curves obtained in the absence of SOC are additionally included
in the spectra (blue dashed lines).
}
\end{figure*}
\begin{figure*}[ht]
\includegraphics[width=\textwidth]{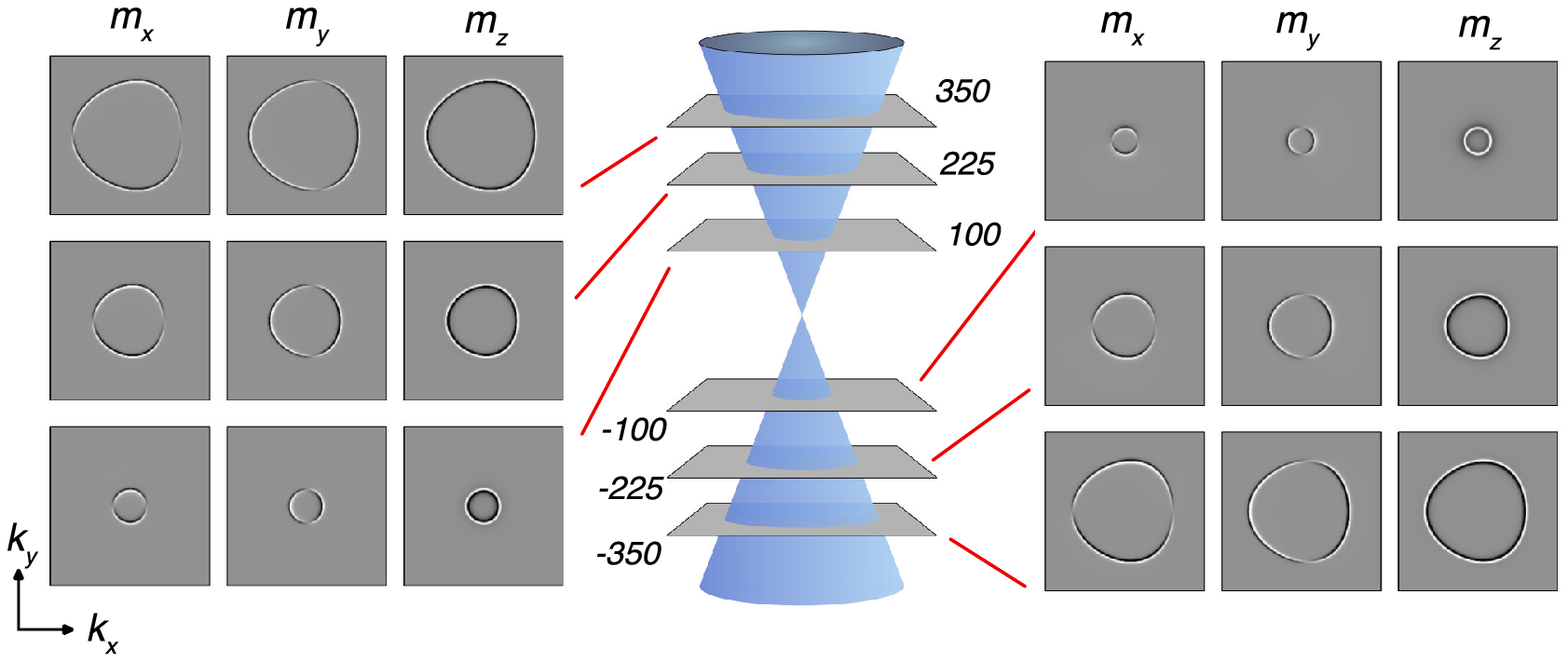}
\caption{\label{fs}
Unfolded G-projected \mkk \ maps calculated at selected values of energy and 
within $k$-regions of the Dirac cones (see sketch in the middle panel with
the energies given in meV). 
Left-hand (right-hand) panel displays the energy cuts at the upper (lower) 
Dirac cones. The $x/y/z$ components of $\vec{m}$ are plotted separately in the
first/second/third column of each panel employing a grey-scale color scheme 
with positive (negative) values of magnetization drawn in white (black).}
\end{figure*}

In this section we will focus on the G/Au/Ni(111) system for which we have
considered two different interface models based on the related STM study of
Jacobsen {\it et al}.\cite{interface}  
The first one consists of graphene pseudomorphically grown on an unreconstructed 
Ni(111) surface with an intercalated Au monolayer between them, which
results in a large moir\'e pattern involving (9$\times$9) and
(8$\times$8) supercells at the G/Ni and Au, respectively 
(see Fig.~\ref{gold}(a)) --in this configuration the Au-Au intralayer distances
are compressed by just 2\% with respect to those in the Au-$fcc$ bulk phase.
The same commensurate lattice applies to the second model, where the Ni top 
layer presents a (9$\times$9) reconstruction after removal of five Ni atoms
followed by a shift of a ten-atom triangle from {\it fcc} to {\it hcp} registry. 
However, all results for this latter model are presented in the Appendix~C
since, to our surprise, the electronic structure of graphene is hardly modified
by such a severe reconstruction. 
Indeed, the
computed adsorption (physisorption) energy is the same for both models (38~meV/C).

After relaxation we found a weak graphene/Au interaction 
with an uncorrugated G layer located at 3.39~\AA\ above the gold layer (see Fig.~\ref{gold}(b)),
slightly larger than the values obtained in previous studies of G/Au interface
(3.2-3.3~\AA \ in Refs.~[\onlinecite{gold-dft, gold-exp, gold-rpa, gold-prb}]),
although an eventual error smaller than 0.2~\AA\ should not alter the 
SOC splittings by more than a few tens of meV.\cite{calleja}
Unexpectedly, 
the Au layer presents a rather small corrugation of 0.08~\AA, similar to
that at the Ni topmost layer (0.15~\AA). 
As regards the magnetic properties of the system,
the Ni substrate's spin polarization was always set along the surface normal 
($+z$ axis) while the intercalated gold was found to couple antiferromagnetically
to it with small induced magnetic moments of 
$\sim -0.02\mu$B/atom versus the $\sim +0.6\mu$B/atom at the Ni surface
layers and the negligible $+0.002\mu$B/atom in graphene.

PDOS\ke\ maps projected on the first atomic layers of the surface are shown in 
Figure~\ref{gold}(d). The contributions of G, Au and Ni are superimposed within
the same map each plotted in a different color (red, light blue and dark blue, 
respectively). Recall that due to the BZ back-folding which transforms both 
$K_G$ and $K'_G$ into $\Gamma$, the two cones appear superimposed in the 
figure.
This time the linear $\pi$-bands (red) are almost 
fully preserved within a $\pm$1~eV interval around $E_F$,
confirming its quasi-freestanding character.
The intense dark blue horizontal bands centered at around
$-0.7$ and $+0.2$~eV correspond to the upper part of the majority and minority
Ni $d$-bands, respectively. Gold $sp$ bands, 
including a surface state with onset at -0.33 eV, analogous to the Au(111) Shockley-type 
$L$-band,\cite{soc, soc_exp}
cross the BZ at both positive and negative energies while
faint traces of the Au $d$-bands (light blue) appear below $-1$~eV inducing
large distortions in the graphene's $\pi$-bands. The decomposition of the maps 
around this region into the diagonal $\uparrow\uparrow$ and $\downarrow\downarrow$ 
stencils (majority and minority spin carriers in the absence of SOC) 
reveals marked differences among them (lower inset in (d)), indicating that
the graphene/gold hybridization is spin dependent due to the
already exchange-splitted Au states. In particular, the $\downarrow\downarrow$
stencil is strongly perturbed around $-1.3$~eV and the $\uparrow\uparrow$
at $-1.7$~eV.

Figure~\ref{gold}(e) displays the unfolded PDOS\ke projected on G along the 
$\Gamma-K_G/K_G'-\Gamma$ lines of its primitive BZ. The unfolding permits
to disentangle the $K_G$ and $K'_G$ DCs and visualize each $\pi$-band
independently thus allowing a direct comparison versus the experimental
ARPES results of Ref.\onlinecite{main_nat}. 
The corresponding magnetization maps together with some single 
PDOS and $\vec m$ spectra computed at selected $k$-points
are shown in panels (f) and (g), respectively. 
Due to the $p3m$ symmetry
holding for the entire surface the PDOS\ke map is 
perfectly symmetric between the $\Gamma-K_G$ and $K'_G-\Gamma$ paths as
they are related by a mirror plane. Most notably, the \mke\ map displays the
same symmetry with a vanishing \mpar\ contribution throughout the entire energy
range implying that the in-plane component of the spin remains 
helical. The fact that $m_z$ does not change sign between the two paths implies
that it does not behave as a pseudo-vector any more but, instead, it is
fixed by the magnetic order that completely dominates the out-of-plane
spin component over the SOC.

In the lower energy region,
the hybridizations with the spin-exchange splitted gold $d$-bands 
induce streaky features opening several gaps with giant spin-splittings larger 
than 100~meV (curves A and B). The structure of the corresponding spectra shows
multiple peaks of different intensity, spin dependent broadenings and
magnetic orientations (always confined within the ($\perp,z$) plane).
This sceneario is therefore reminiscent of the strong SOC-mediated interaction 
found between
the G and the Pt-$d$ bands in the previous section.
In the $\pm1$~eV range around $E_F$ (curves C, D, E and F),
the lack of hybridizations leads to quasi-linear sharp bands which even 
allow to resolve the double peak structure in the PDOS$(E)$ in certain cases 
(D or F). However, the splittings never exceed $\sim 10$~meV. 
Curve E shows the magnetization at the gapless Dirac point where
only $m_z$ is non zero --i.e. consistent with a weak intrinsic SOC picture. 
Interestingly, there is a large asymmetry
between the two components, the $+m_z$ being much sharper than the $-m_z$.

In order to try to quantify the interplay between SOC and magnetic exchange
we have included in the spectra of panel (g) the magnetization in the absence 
of the former (blue dashed lines). The similarities with the $m_z$ curves (solid
blue lines) in all the graphs, even in the distorted regions (curves A or B),
reveals that the SOC manifests in this system only as a rather small
perturbation to the electronic structure, its main effect being the emergence
of an in-plane helical component, \mper, in the spin texture (solid red lines).

Fig.~\ref{fs} presents constant energy $\vec{m}(\vec{k})$ surfaces
computed around $K_G$ at several energies above and below the Dirac point,
as sketched at the central panel. The circular features become more trigonally
distorted as the energy is further away from the Dirac point. 
The in-plane projections $m_{x/y}$ make patent the helical character of the 
$\pi$-states as well as a small RS 
since the DCs appear $k$-shifted with respect to each other (as expected, the 
shifts are along the same directions in the upper and lower cones). On the other
hand, the $m_z$ maps do not follow a standard exchange magnetic picture whereby
the entire band structure of each spin branch would be shifted in energy 
(vertically) with respect to the other one. Here, the upper cone couples
ferromagnetically (dark/minority circles lie inside the white/majority 
ones) and the lower cone anti-ferromagnetically to the Au/Ni substrate.
The only exception is the $-100$~meV map where the broad minority band
appears at both sides of the sharper majority circle.

Last, we discuss on the agreement between our results and the 
experimental data obtained via ARPES and SP-ARPES reported in 
Refs.\onlinecite{first13mev, main_nat}. Our calculations reproduce well
the quasi-free-standing character of the G on this surface, as the DCs
appear rather intact  (within the energy interval considered in our work), as
well as the negligible doping and the absence of a gap at the Dirac point. 
Furthermore, the kink observed at $-0.95$~eV in Ref.\onlinecite{first13mev} and
ascribed to an electron-mass renormalization in the DC's dispersion matches
well with the onset of strong perturbations in the $\pi$-bands which in our
simulations appear below $-1$~eV. In fact, if a straight line taken along the 
upper DC in Fig.~\ref{gold}(e) is extrapolated towards negative energies
(not shown), deviations from the lower cone appear around this onset and in 
the same direction as the experiments (towards a smaller electron mass).
Finally, our calculated splitting values (of around 10~meV) perfectly agree 
with those derived in the first SP-ARPES study ($13\pm3$~meV), although we
attribute their origin to the magnetic order rather than to a Rashba-type SOC
which, anyhow, is observable in our simulations. On the other hand, our
results cannot explain the giant ($\sim 100$~meV) splittings measured in
Ref.\onlinecite{main_nat} for one of the two phases that co-existed within
the same surface and attributed to a possible corrugation of the G layer which 
could lead, locally, to unusually small G/Au distances.
Our simulations employing a realistically large supercell seem to
exclude this possibility since the G is found to be only weakly physisorbed and
remains hardly corrugated even after the large surface reconstruction
described in the Appendix~C. A possible alternative explanation could be the
formation of a Ni/Au alloy at the surface layer which could bring the G closer
to the surface due to the stronger G-Ni coupling, as has been recently proved 
for the G/Fe/Ir(111) system.~\cite{2dmaterials} Still, and based on the
results presented here, we believe that giant splittings will not show
up in the quasi-linear regions but only at anti-crossings (gaps) that could 
emerge close to $E_F$ due to the presence of Ni atoms at the top layer.\cite{g-ni}

\section{Summary and conclusions}
In summary, we have unveiled with unprecedented detail the SOC-induced spin 
texture in the graphene's $\pi$ bands arising from the presence of a heavy metal
semi-infinite substrate.  
We have considered two paradigmatic systems:
a non-magnetic Pt(111) and a spin-polarized Au/Ni(111) surface.
For the former we find that the SOC splitted continuum of metal bands 
hybridize with the Dirac cones at multiple energies and manifest
as local spin-dependent deformations in the linear bands at the avoided
crossing gaps,
often exceeding 100~meV, accompanied by spin reorientations at
the edges, while in between gaps (quasi-linear regions) a reduced splitting
of at most 10-20 meV remains. Both the precise value of the splitting as well 
as the G's induced magnetization change continuously as the splitted branches
disperse, typically crossing between them and bearing different widths.
Therefore, we conclude that the giant SO splittings reported for this system 
are most probably located close to anti-crossing gaps.
The number and energy location of the avoided-crossings
will in general depend on the particular moir\'e pattern since, as the
metal BZ shrinks or rotates with respect to that of graphene, different metal
bands will mix (see the
comparison between the (2$\times$2)- and (3$\times$3)-G/Pt phases in Fig.~\ref{fold}).
Obviously, as the supercell (moir\'e pattern) increases
in size further crossings will be expected leading to even more complex
spin textures.

In the case of G/Au/Ni(111), the absence of gold $d$-states around the Fermi 
level leaves continuous linear $\pi$-branches with small splittings in the 
10~ meV range. However, their origin is mainly the substrate's magnetic order
that is transferred to graphene, while the induced SOC introduces a
helical in-plane component to the magnetization but hardly alters the G's
total PDOS. 
No evidence for large RSs of the Dirac cones in this energy region
was found for either of the two different phases considered, at contrast with 
a previous experimental study;~\cite{main_nat} we assign this
discrepancy to a possible Au/Ni alloying process at the surface layer.
The presence of Au-$d$ states below $-1$~eV, on the other hand, leads again to 
a complex spin structure similar to that found for both G/Pt phases.

We note that unveiling such a complex spin texture represents
a key finding which 
not only opens a new challenge to current (SP-)ARPES based studies of the SO proximity
effect, but should also stimulate further research aiming at engineering the {\it relativistic} 
electron's spin. Indeed, there are several non-negligible problems to be solved. 
On one hand, splitting the two spin branches by sizeable energies
($> 100$~meV) seems only possible at anti-crossing regions with the metal's
$d$-states where the $\pi$-bands lose their linear character as gaps are
opened and therefore electronic transport across the G layer is suppressed.
Furthermore, no spin-momentum locking holds in these regions as the G's
induced magnetization acquires very different orientations in space depending
on the specific metal band with which it hybridizes. On the other hand, at
quasi-linear regions, either in between these gaps or where no $d$-bands
are present (typically the upper DC), only very small splittings ($< kT$)
can be observed, with the further disadvantage that
the two branches do not necessarily disperse strictly parallel and may even
cross yielding an interchange of magnetizations.
The broadening induced by the substrate's self-energy is also found to be 
highly relevant in the G's final spin texture, as it strongly varies
between the two spin components often leading to single sharp peaks emerging 
from a broader background signal. 

Our results show a trade-off between large SOC-derived splittings and the
loss of linearity of the $\pi$ bands, which needs to be overcome in order 
to harness the SOC-induced magnetization. 
We believe, however, 
that our detailed analysis of the spin-orbit proximity effect in graphene adsorbed
on metallic substrates, far beyond the resolution of current photoemission
techniques, will not only motivate further experimental studies aiming to capture
the G's spin properties, but will also facilitate their correct interpretation. Moreover, it remains to address any topological properties that could emerge at the multiple gaps opened. Intrinsic SOC transferred to the
G via the proximity effect is known to induce a QSHE state after small gap 
openings (of at most a few tens of meV) at the Dirac point~\cite{weeks} while 
here much large gaps ($100-200$~meV) have been systematically found away from 
$K_G$ often crossed by a non-vanishing density of states.

\section{Acknowledgments}
We are deeply grateful to Prof. A. Arnau for the critical reading of the
manuscript and his highly stimulating comments.
J.S. acknowledges Polish Ministry of Science and Higher Education for financing
the postdoctoral stay at the ICMM-CSIC in the frame of the program Mobility 
Plus. J.I.C acknowledges support from the Spanish Ministry of
Economy and Competitiveness under contract No. MAT2015--66888-C3-1R.

\appendix
\section{Brillouin Zone unfolding}
\begin{figure}[b]
\includegraphics[width=0.50\textwidth]{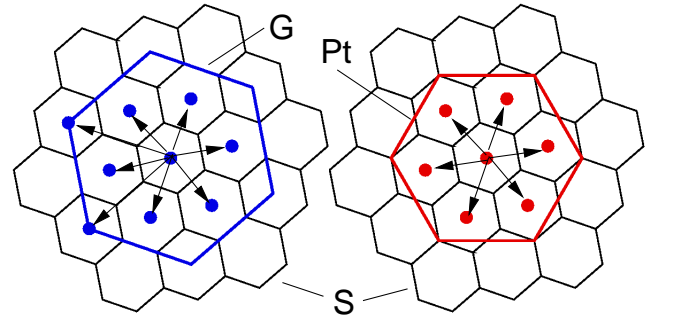}
\caption{\label{bzs} BZ schemes for the G-(3$\times$3) lattice 
(blue hexagon in left-hand figure), the Pt-$(\sqrt{7}\times\sqrt{7})$R19.1$^\circ$
(red in right-hand figure), and their commensurate cell $S$ (black hexagons in both
figures). The $\vec G$-vectors pointing to the $S$ BZs enclosed by the G 
and Pt BZs (9 and 7, respectively) are indicated by the arrows 
(apart from the $\Gamma$ point $\vec G=0$).}
\end{figure}

In the case of moir\'e patterns (or supercells in general), where lattices at 
layers $I$ and $J$ are different and a common supercell $S$ exists for both, 
any $k$-point within the BZ of layer $I$, $\vec k_I$, may be expressed as 
$\vec k_I=\vec k_S + \vec G_I$, where $\vec k_S$ is confined within the BZ of
$S$ and $\vec G_I$ are the so-called $\vec G$-vectors\cite{green} that relate 
the reciprocal lattices of $S$ and $I$; the number of these vectors is given by
the ratio between the unit cell areas of $S$ and $I$, $N_{SI}$ (see Fig.~\ref{bzs}).
Since for the combined system only $\vec k_S$ is conserved, any quantity in 
$k$-space projected at layer $I$ will mix all $\vec G_I$ vectors (BZ folding).
However, when the $I$-$J$ interaction is weak, one may expect that, to
a good approximation, translation symmetry is preserved at layer $I$ and
hence, $\vec k_I$ is approximately conserved within this layer. We explain
below in detail our procedure to obtain such unfolded band structure
from a supercell slab calculation. 

Considering the general case where the substrate's unit cell is
the same as that of the supercell $S$ (i.e. a reconstructed surface) and
denoting by the index $I$ the graphene layer adsorbed on top whose electronic 
structure we wish to unfold,
we first construct the intra-layer Hamiltonian in $k$-space as:
\begin{equation} 
  H_{II}(\vec k_I)= \frac{1}{N_{SI}} \sum_{\vec R_I,\vec L_I} e^{i \vec k_I (\vec R_I-\vec L_I)} H_{II}(\vec R_I-\vec L_I)
\end{equation}
\noindent
where $\vec R_I$ runs over the lattice vectors at $I$, $\vec L_I$ are the
vectors linking the origin of the layer with that of each of the $N_{SI}$ unit 
cells contained in the $S$ supercell, and $H_{II}(\vec R_I-\vec L_I)$ contains 
the matrix elements between the G's unit cell centered at $\vec L_I$ and those
shifted by $\vec R_I$ --all of them available from the self-consistent slab's
Hamiltonian previously computed and saved. Equation (A1) thus averages the 
matrix elements between the carbon A and B atoms over the entire supercell so 
that traslational symmetry is {\it imposed} within the layer.
Next, we express the $k$-space interactions between $I$ and $S$ as:
\begin{equation}
  H_{IS}(\vec k_S,\vec G_I)= \frac{1}{\sqrt{N_{SI}}} \sum_{\vec R_S,\vec L_I} e^{i (\vec k_S+\vec G_I) (\vec R_S-\vec L_I)} H_{IS}(\vec R_S-\vec L_I)
\end{equation}
\noindent
where, again, all inter-layer $H_{IS}(\vec R_S-\vec L_I)$ matrix
elements can be extracted
from the self-consistent supercell slab calculation. The Green's
function projected at layer $I$ may then be expressed according to:
\begin{equation}
  G_{II}(\vec k_S,\vec G_I,E)= \left( F_{II}(\vec k_I,E)+\Sigma_{II}^S(\vec k_S,\vec G_I,E) \right)^{-1}
\end{equation}
\noindent
where $F=E\cdot O - H$ is the secular matrix and $\Sigma_{II}^S(\vec k_S,\vec G_I)$
stands for the self-energy at layer $I$ arising from the presence of the
subtrate, which may be calculated via Green's functions techniques following the
same $k$-scheme as for $H_{IS}(\vec k_S,\vec G_I)$.\cite{loit}
Unfolding the BZ at $I$ may now be easily carried out by extracting the 
contribution of each $\vec k_I=\vec k_S+\vec G_I$ term individually. 
In practice, we run $\vec k_I$ along a given direction
and only retain the $\vec G_I=0$ contribution of $G_{II}(\vec k_S,\vec G_I,E)$.

Similar to the case of G, BZ unfolding may also be performed for any
substrate layer as long as it is not strongly reconstructed. 
The above procedure remains valid, although for the surface layer
we need to consider two inter-layer interactions (self-energies) instead
of just one: that involving 
the G and that with the rest of the substrate below. In general, to reduce 
the number of approximations associated to this approach we perform the 
unfolding for the G and the surface metal layer in separate calculations.

Let us finally note that our unfolding scheme is exact as long as translational
symmetry is strictly preserved at layer $I$. 
For the quasi-freestanding G case, either on Pt(111) or Au/Ni(111), the G-metal
interactions are indeed weak and it is clearly a good approximation.

\section{G/Pt(111) folded band structures}
\begin{figure}[ht]
\includegraphics[width=0.5\textwidth]{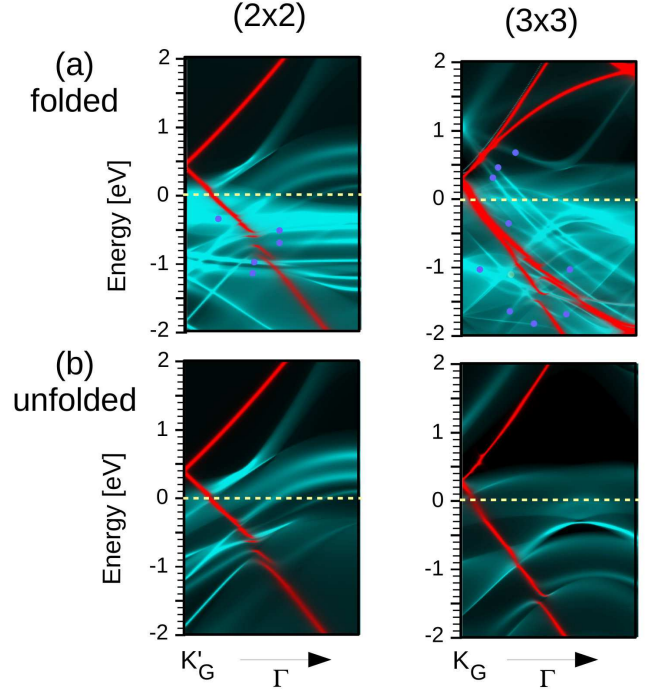}
\caption{\label{fold} DOS\ke\ maps projected on the G (red) and first Pt
layer (light blue) calculated for the G/Pt (2$\times$2) 
(left column) and (3$\times$3) (right column) phases along the B $k$-paths 
shown in Fig.~\ref{geom}.
(a) Unfolded band structure after assuming that the (1$\times$1) translational
symmetry is preseved at the G and surface Pt layers (see Appendix~A) and, 
(b) folded band structure employing the (2$\times$2)/\retwo\ and 
(3$\times$3)\rethree\
supercells to describe all G/Pt layers.
}
\end{figure}

In Figure~\ref{fold} we present a comparison of the folded (top panels)
versus unfolded (bottom panels) PDOS\ke\ maps for both the G/Pt (2$\times$2) and
(3$\times$3) phases computed along the $K'_G-\Gamma$ direction ($k$-paths B
in Fig.~\ref{geom}(b)).
We have simultaneously merged in each map the G (colored in red)
and first Pt layer (light blue) projections. For the (2$\times$2) phase
(left column) around five new back-folded Pt bands appear below $E_F$ and 
cross the G's $\pi$-band (each marked by a small blue circle),
while the upper cone remains identical to that of the unfolded case due to
the absence of extra bands. The profusion of back-folded Pt bands is much larger
in the (3$\times$3) phase, as should be expected due to its smaller BZ size. 
In this geometry both DCs ($K_G$ and $K'_G$) are back-folded into the $\Gamma$
point and hence, the two branches appear superimposed in the upper map.
We have
again marked with small circles those which do not appear in the unfolded
map (up to eight clearly visible) with the peculiarity that this time, two of 
them cross the upper DC inducing the small distortions around 0.6~eV that can 
be appreciated in the unfolded map below.

\section{G/Au/Ni(111) reconstructed model}
\begin{figure*}[ht]
\includegraphics{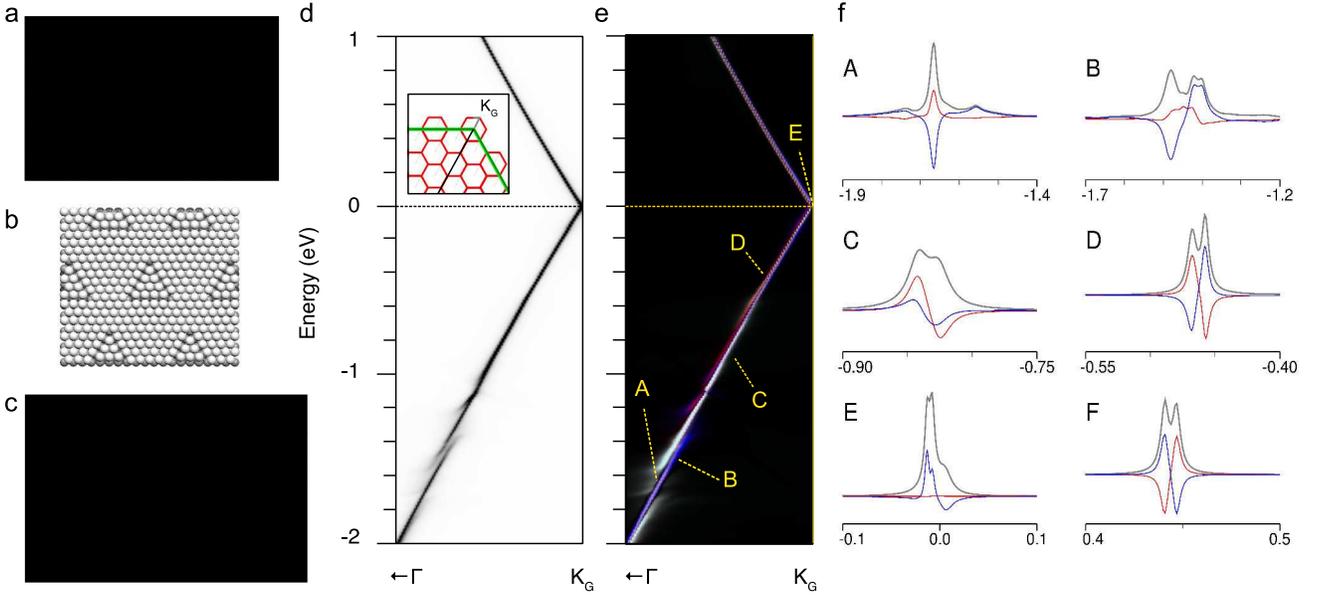}
\caption{\label{nir} (a-c) Top and side views of the relaxed geometry for
the reconstructed G/Au/Ni(111) configuration. C, Au and Ni atoms are drawn
as red, gold and gray balls, respectively. In (b) the Au and G layers have
been removed in order to show the Ni triangular motif characteristic of
this reconstruction. In (c) first interlayer average 
spacings (black font) and intra-layer corrugations (blue font) are indicated 
in \AA.
(d) Electronic structure and (e) corresponding magnetization density unfolded 
and projected on graphene along the $\Gamma-K_{G}$ line of its primitive BZ 
(shown in the inset). Color scheme same as in the maps shown in the main 
manuscript. (f) Single spectra extracted from panels (d) and (e) at the
$k$-points marked in the latter; PDOS$(E)$ in solid gray and 
$m_{\perp/z}(E)$ components in red/blue (the $m_{\parallel}$ is zero along this $k$-line).}
\end{figure*}

We describe here the results for an alternative G/Au/Ni(111) 
model, whose geometry differs from the one considered in the main text by the Ni substrate's reconstruction, based on previous STM studies~\cite{interface} 
where the deposition of Au on Ni(111) at RT yielded a complex 
$\sim(9.7\times9.7$) reconstruction with triangular motifs assigned to a 
strongly reconstructed top Ni layer whereby five vacancies are created and ten
surface Ni atoms forming a triangle are shifted from an $fcc$ to an $hcp$ 
registry (see Figure~\ref{nir}(b)).~\cite{interface, interface2}
Although there is no experimental evidence for the existence of such 
configuration when graphene is deposited on top, we have considered it as a 
possible explanation for the appearance of two phases with different splittings
as reported in the SP-ARPES experiments~\cite{main_nat}.

In Fig.~\ref{nir}(a-c) we summarize the relaxed geometry of such configuration;
for simplicity, we have assumed the same G/Au/Ni  
$(9\times9$)/(8$\times8$)/($9\times9)$ commensurate supercell as for the
unreconstructed model since it can nicely accommodate the large triangular
motifs. Although the triangular Ni motifs induce a 
large buckling in the Au layer of 0.7~\AA, the corrugation of graphene remains 
negligible (0.03~\AA) with an average G-Au distance of $d_{G-M}=3.48$~\AA\ 
only slightly larger than in the unreconstructed model considered in the main text. 
The Au layer again exhibits a small magnetic moment of 0.03 $\mu$B/atom 
antiferromagnetically coupled to that of the Ni surface. 
Figure~\ref{nir}(d-f) presents the unfolded PDOS\ke and magnetization maps together with single spectra extracted from the latter, following the same scheme as in Fig.~\ref{gold}. The maps corresponding to the unreconstructed and reconstructed phases are qualitatively very similar, with an
almost intact (only weakly spin-splitted) upper Dirac cone and 
the lower cone strongly textured due to multiple hybridizations with the
metal bands particularly below $-1$~eV. 
At the Dirac point (spectra E), two overlapping peaks can be seen signaling
a larger intrinsic SOC for this model, although their broadenings, 
different for each spin, are both sufficiently large to close any gap.
Nevertheless, and as can be seen from the single spectra in panel (f), the 
magnitude of the spin-splittings along the DC do not exceed a few tens of meV 
in the quasi-linear regions. Therefore, and similar to the unreconstructed model
described in the main text, the triangular surface reconstruction
cannot explain either the giant RS splitting phase recently reported for this 
system~\cite{main_nat}.

\end{document}